\documentclass{trbunofficial}
\usepackage{graphicx}
\usepackage{xcolor}
\usepackage{booktabs}
\usepackage{threeparttable}

\usepackage[hidelinks]{hyperref}

\AuthorHeaders{Saba, Dasgupta, Rahman, Huynh, Zhao, Vuran, Liu, and Ozguven}
\title{Infrastructure-Enabled Risk Assessment of Hazardous Road Conditions on Rural Roads during Inclement Weather} 
\author{
  \textbf{Suhala Rabab Saba}\\
  Department of Civil, Construction \& Environmental Engineering, The University of Alabama\\
  Smart Communities and Innovation Building (SCIB), 28 Kirkbride Lane,\\
  Tuscaloosa, AL 35487-0288\\
  Email: ssaba@crimson.ua.edu \\
  \hfill\break
  \textbf{Sagar Dasgupta}\\
  Department of Civil, Construction \& Environmental Engineering, The University of Alabama\\
  Smart Communities and Innovation Building (SCIB), 28 Kirkbride Lane, 
  Tuscaloosa, AL 35487-0288\\
  Email: sdasgupta@ua.edu \\
  \hfill\break%
  \textbf{Mizanur Rahman}\\
  Department of Civil, Construction \& Environmental Engineering, The University of Alabama\\
  Smart Communities and Innovation Building (SCIB), 28 Kirkbride Lane, 
  Tuscaloosa, AL 35487-0288\\
  Email: mizan.rahman@ua.edu \\
  \hfill\break
  \textbf {Nathan Huynh}\\
  Department of Civil and Environmental Engineering, University of Nebraska-Lincoln\\
  262D Prem Paul Research Center at Whittier School\\
  2200 Vine Street, Lincoln 68588 Nebraska\\
  Email: nathan.huynh@unl.edu\\
  \hfill\break%
  \textbf {Li Zhao}\\
  Department of Civil and Environmental Engineering, University of Nebraska-Lincoln\\
  262K Prem Paul Research Center at Whittier School\\
  2200 Vine Street, Lincoln 68588 Nebraska\\
  Email: lizhao@unl.edu, (402) 472-1928\\
  \hfill\break%
  \textbf {Mehmet C.~Vuran}\\
  School of Computing, University of Nebraska-Lincoln\\
  SHOR 214, Lincoln, NE 68588-0150 a\\
  Email: mcv@unl.edu\\
  \hfill\break%
  \textbf {Qiang Liu}\\
  School of Computing, University of Nebraska-Lincoln\\
  AVH 262, Lincoln NE 68588-0115\\
  Email: qiang.liu@unl.edu\\
  \hfill\break%
  \textbf {Eren Erman Ozguven}\\
  Department of Civil and Environmental Engineering, FAMU-FSU College of Engineering\\
  2525 Pottsdamer Street, Tallahassee, FL, 32311\\
  Email: eozguven@eng.famu.fsu.edu\\
  \hfill\break%
}

\begin{document}
\maketitle

\section{Abstract}

Rural roadways often expose Commercial Motor Vehicle (CMV) drivers to hazardous conditions, such as heavy fog, rain, snow, black ice, and flash floods, many of which remain unreported in real time. This lack of timely information, coupled with limited infrastructure in rural areas, significantly increases the risk of crashes. Although various sensing technologies exist to monitor individual hazards like low visibility or surface friction, they rarely assess the combined driving risk posed by multiple simultaneous hazards, nor do they provide actionable recommendations such as safe advisory speeds. To address this critical gap, in this study, we present a roadway hazard risk assessment framework that provides an approach to quantify the probability and severity of crash occurrences due to specific roadway hazards. To evaluate this framework, we presented a case study by constructing a synthetic "year-long" dataset that encompasses every possible pairing of road surface and visibility conditions.
Our analysis confirms that the combined Probability-Severity scoring yields a coherent, stepwise risk profile across all hazard scenarios. These results validate the practicality of our risk assessment approach and provide a foundation for deploying graduated safety measures in real-world roadway operations.

\hfill\break%
\noindent\textit{Keywords}: driving risk assessment, safe advisory speed, road surface friction, poor visibility, hazardous road conditions, commercial motor vehicles, and rural roadways.
\newpage

\section{Background and Motivation}

Despite comprising only 19 percent of the U.S. population, rural America bears a disproportionate burden of roadway fatalities. Between 2016 and 2020, 45 percent of all roadway deaths occurred on rural roads, with the fatality rate in these areas being twice as high as that on urban roadways \cite{raymond2022america}. According to the 2022 America’s Rural Roads report by the Governors Highway Safety Association (GHSA), 84 percent of rural crashes occur away from intersections, indicating the prevalence of mid-block incidents, often in poorly monitored, remote segments \cite{raymond2022america}. The 2023 National Highway Traffic Safety Administration (NHTSA) report \cite{nhtsa2023trafficsafety} further underscores the gravity of rural road safety challenges, highlighting that roadway departure and head-on collisions are the two most common fatal crash types in rural areas. Critically, over half (55 percent) of these fatal rural crashes involve large trucks, and in 62.3 percent of these cases, the trucks were simply traveling straight a reflection of the hazardous environment rather than driver maneuvering \cite{nhtsa2023trafficsafety}. Additional insights reveal that crash severity escalates during specific temporal and environmental conditions. Weekend crashes, particularly those occurring between 9 p.m. and 6 a.m., are associated with reduced visibility, increased wildlife movement, and driver fatigue factors that disproportionately affect Commercial Motor Vehicle (CMV) operators who often traverse these roads during off-peak hours. Adverse weather conditions such as snow, ice, and rain further compound these risks. According to the 2020 Federal Motor Carrier Safety Administration (FMCSA) crash facts, 3,048 truck-involved crashes occurred on icy or frosted roads, and 13 percent of all truck crashes happened during inclement weather \cite{FMCSA2022_LTB_CF2020,nhtsa2023trafficsafety}. Strikingly, icy road conditions alone contribute to more fatalities in the United States than all other weather events combined, including hurricanes, tornadoes, and severe storms \cite{FMCSA2022_LTB_CF2020}. These statistics emphasize the pressing need for real-time risk assessment and advisory systems tailored to rural roadway conditions.

In practice, roadway hazards take numerous forms: extremely poor visibility due to fog, snow, or heavy rain; low surface friction from black ice or water films; hydroplaning risks; sudden stops due to stalled vehicles; and flash flooding. Unfortunately, such conditions often remain unreported in real time, undermining the ability of drivers especially CMV operators to anticipate and respond to dangerous roadway scenarios. This limitation is particularly severe on rural roads, where monitoring infrastructure is often sparse or altogether absent due to the prohibitive cost and logistical complexity of deploying and maintaining advanced sensing systems in remote areas. As a result, CMV drivers navigating rural routes face a significant safety disadvantage, often relying on personal judgment in rapidly changing and unpredictable driving environments.

Although there are existing technologies capable of monitoring individual road conditions, such as visibility sensors, black ice detectors, and rain gauges, they are typically designed to sense and report specific, isolated hazards. For instance, visibility sensors can assess fog density, and optical ice detectors can identify black ice patches. Roadside edge computing systems can process this sensor data in real time. However, a critical limitation of these current solutions lies in their siloed nature; they lack the capability to assess the compound driving risk associated with multiple concurrent hazards and do not provide actionable guidance such as safe advisory speeds tailored to specific road and weather conditions.

Multiple research efforts have attempted to assess driving risk by modeling driver behavior under adverse conditions. These models typically consider visibility range, surface friction, and time headway as inputs for estimating crash likelihood \cite{chen2019influence,das2017safety,santos2021relating,yasanthi2021modelling,chen2019vehicle,li2025car}. For example, the safe stopping distance model incorporates visibility and friction to calculate the minimum required distance between vehicles to avoid collisions. However, these models often rely on assumptions that do not hold under the combined influence of multiple simultaneous hazards. Speed advisory systems have also been explored in literature, particularly in the context of weather-responsive traffic management (WRTM). Study by FHWA \cite{Katz2012_VSL_WetWeather} emphasized the use of real-time friction and visibility data to generate variable speed limits on highways. While effective in controlled environments, these systems are not widely deployed on rural roads and lack the adaptive capabilities to address compound hazards. Despite substantial progress in sensor technologies and risk modeling, several critical gaps remain. First, most existing systems are designed to detect and respond to individual hazards and fail to account for the interactive effects of multiple concurrent hazards.  Third, current approaches do not provide actionable and interpretable guidance to drivers on multiple hazard severities.

To address this critical gap, in this study, we present a roadway multi-hazard risk assessment framework that provides an approach to quantify the probability and severity of crash occurrences due to specific roadway hazards. To quantify the probability of a crash occurrence, we consider road surface friction and visibility under prevailing weather conditions. On the other hand, we quantify the severity of a roadway hazard by determining the safe advisory speed considering road surface condition and sight distance due to inclement weather and the corresponding percentage reduction from the roadway’s posted speed limit. To evaluate this framework, we presented a case study by constructing a synthetic “year-long” dataset that encompasses every possible pairing of road surface and visibility conditions.

\section{RISK ASSESSMENT APPROACH OF ROADWAY HAZARDS}

\begin{figure}[!ht]
  \centering
  \includegraphics[width= 1\textwidth]{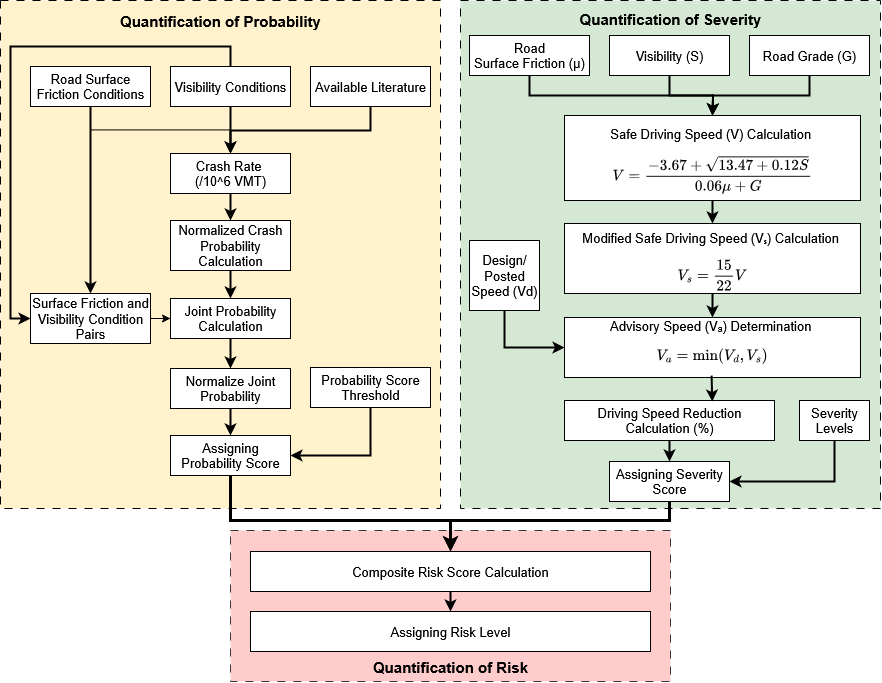}
  \caption{ Risk quantification framework for roadway hazards}
  \label{flowchart}
\end{figure}

Risk Assessment is the systematic process of identifying, quantifying, and evaluating any potential hazards. \ref{eq:risk} has been used to quantify the risk considering the probability and severity of any hazard \cite{kanj2025agent}. However, it is necessary to quantify the probability and severity of any hazards depending on their nature and applicability. We can define roadway hazards in many different forms: extremely poor visibility due to fog, snow, or heavy rain, and low surface friction from black ice, water films, and flash flooding. In this study, we present a roadway hazard risk assessment framework that provides an approach to quantify the probability and severity of crash occurrences due to specific roadway hazards. To quantify the probability of a crash occurrence, we consider road surface friction and visibility under prevailing weather conditions. On the other hand, we quantify the severity of a roadway hazard by determining the safe advisory speed considering road surface condition and sight distance due to inclement weather and the corresponding percentage reduction from the roadway’s posted speed limit. Figure \ref{flowchart} illustrates our proposed methodology. 

\begin{equation}
\mathrm{Risk} \;=\; \mathrm{Probability} \times \mathrm{Severity}
\label{eq:risk}
\end{equation}

In the following subsections, we first detail the probability quantification approach, which converts published crash‐rate data into normalized marginal and joint probabilities before assigning discrete probability Scores.  We then describe the severity quantification workflow, in which safe advisory speeds are computed from surface friction, sight distance, and roadway geometry, and translated into severity scores based on percentage speed reductions. Finally, we explain the risk quantification approach using probability and severity to yield a composite risk score and corresponding risk‐level category. 

\subsection{Quantification of Probability}
Table \ref{tab:marginal_probs} provides crash rates (crashes per $10^6$ VMT) in terms of road surface condition and visibility range from transportation‐safety literature \cite{McCann2016_VTRC17R4,Qiu2009_PerfHighwayWinterMaint}. We observe that road surface friction coefficient values reduce as the crash rate value increases. This trend is also evident in the visibility data: as visibility distance decreases, the crash rate increases. Reduced visibility impairs drivers’ ability to detect and respond to hazards in a timely manner, thereby elevating the risk of collisions. Both of these findings indicate that the reduction of surface grip between tire and road, due to the friction coefficient value and reduced visibility, are the key drivers of elevated crash probability. Both of these factors are related to the weather, which allows for their simultaneous occurrence, thereby increasing the overall crash risk associated with them. Thus, using the findings from the literature, we leveraged the four marginal categories of surface friction (i.e., dry, wet, snow, and icy) and four visibility bands (i.e., clear, rain/snow, dense fog, very dense fog) to enumerate all realistic environmental scenarios. By pairing each friction regime with each visibility condition, we generated 16 unique driving-environment scenarios as shown in the table \ref{tab:complete_combinations}.

\begin{table}[ht]
\centering
\caption{Realistic Scenarios Combining Friction and Visibility}
\label{tab:complete_combinations}
\begin{tabular}{cllc}
\toprule
\# & \textbf{Friction Condition}    & \textbf{Visibility Condition}         & \textbf{Practicality}                         \\
\midrule
1  & Dry ($\mu$: 0.7–0.9)   & Clear (> 1640 ft)         & Common (normal driving)                             \\
2  & Dry ($\mu$: 0.7–0.9)   & Rain/Snow (328–656 ft)    & Rare (brief post-rain dry roads)                   \\
3  & Dry ($\mu$: 0.7–0.9)   & Dense Fog (164–328 ft)    & Possible (radiation fog on dry pavement)           \\
4  & Dry ($\mu$: 0.7–0.9)   & Very Dense Fog (< 164 ft) & Very Rare (extreme fog, no residual moisture)      \\
5  & Wet ($\mu$: 0.4–0.6)   & Clear (> 1640 ft)         & Common (roads slowly drying after rain)            \\
6  & Wet ($\mu$: 0.4–0.6)   & Rain/Snow (328–656 ft)    & Common (ongoing precipitation)                     \\
7  & Wet ($\mu$: 0.4–0.6)   & Dense Fog (164–328 ft)    & Possible (humid/fog during or after rain)          \\
8  & Wet ($\mu$: 0.4–0.6)   & Very Dense Fog (< 164 ft) & Uncommon (heavy fog while wet)                     \\
9  & Snow ($\mu$: 0.2–0.3)  & Clear (> 1640 ft)         & Common (post-snowfall clear skies)                 \\
10 & Snow ($\mu$: 0.2–0.3)  & Rain/Snow (328–656 ft)    & Rare (mixed sleet/rain over snow)                 \\
11 & Snow ($\mu$: 0.2–0.3)  & Dense Fog (164–328 ft)    & Rare (cold fog over snow-laden roads)              \\
12 & Snow ($\mu$: 0.2–0.3)  & Very Dense Fog (< 164 ft) & Common (active snowfall with low visibility)       \\
13 & Icy ($\mu$: 0.05–0.15) & Clear (> 1640 ft)         & Possible (morning black ice before melting)        \\
14 & Icy ($\mu$: 0.05–0.15) & Rain/Snow (328–656 ft)    & Rare (freezing rain conditions)                    \\
15 & Icy ($\mu$: 0.05–0.15) & Dense Fog (164–328 ft)    & Rare (ice fog in extreme cold)                     \\
16 & Icy ($\mu$: 0.05–0.15) & Very Dense Fog (< 164 ft) & Common (snow/ice with blowing snow)  \\
\bottomrule
\end{tabular}
\end{table}

Next, we converted the crash rate values for individual conditions of both friction and visibility into relative probability values, representing the likelihood of a crash occurring under each specific condition, as shown in Table \ref{tab:marginal_probs}. To achieve this, we normalized each set of empirical crash rates independently, transforming them into valid probability distributions.

\begin{equation}
P_{f}(i) \;=\; \frac{R_{f}(i)}{\sum_{j}R_{f}(j)},
\qquad
P_{v}(k) \;=\; \frac{R_{v}(k)}{\sum_{\ell}R_{v}(\ell)}
\label{eq2}
\end{equation}\\

Equation \ref{eq2} rescales raw crash‐rate values \(R\) into valid probability distributions. For each friction category \(i\), $P_{f}(i)$, is obtained by dividing its crash rate $R_{f}(i)$ by the sum of all friction crash rates $\sum_{j} R_{f}(j)$. Similarly, for each visibility band \(k\), $P_{v}(k)$, is the visibility crash rate $R_{v}(k)$ over the total visibility crash rates $\sum_{\ell} R_{v}(\ell)$. This normalization ensures that the resulting probabilities across each marginal set sum to one, allowing direct comparison and combination in the joint probability computation. The overall findings are shown in Table \ref{tab:marginal_probs}. 

\begin{table}[ht]
\centering
\caption{Marginal Crash Rates and Normalized Probabilities}
\label{tab:marginal_probs}
\begin{tabular}{lcc}
\toprule
\textbf{Condition} & \textbf{Crash Rate (per 10$^{6}$ VMT)} & \textbf{Normalized Probability} \\
\midrule
\multicolumn{3}{l}{\textbf{Friction}} \\
Dry ($\mu$ : 0.7 - 0.9)    & 1.90  & 0.0943 \\
Wet  ($\mu$ : 0.4 - 0.6)  & 3.75  & 0.1862 \\
Snow ($\mu$ : 0.2 - 0.3)   & 5.50  & 0.2730 \\
Icy ($\mu$ : 0.05 - 0.15)   & 9.00  & 0.4466 \\
\addlinespace
\multicolumn{3}{l}{\textbf{Visibility}} \\
Clear (>\,1640 ft)   & 0.685 & 0.0262 \\
Rain/Snow (328–656 ft)   & 1.85  & 0.0706 \\
Dense Fog (164–328 ft)   & 4.95  & 0.1890 \\
Very Dense Fog (<\,164 ft)    & 18.70 & 0.7142 \\
\bottomrule
\end{tabular}
\begin{tablenotes}
\item $\mu$ = road surface friction coefficient
\end{tablenotes}
\end{table}

To capture how combined road‐surface and visibility conditions influence crash likelihood, we compute a joint probability for each of the 16 friction–visibility pairs as from table \ref{tab:complete_combinations}. We assume that the effects of pavement friction and sight distance on crash occurrence act independently. In practice, this means a wet road does not change how fog affects visibility probability, and vice versa. While surface and visibility can interact in reality, treating them as independent simplifies the model. Using the empirical equation of finding the joint probability value of two independent events, the raw joint probability of a crash under both conditions can be expressed as,
\\
\begin{equation}
P_{ij} \;=\; P_{f}(i)\;\times\;P_{v}(j)
\end{equation}\\

Renormalization was done to the newly obtained values for each of the 16 friction–viscosity pairs to restore a proper probability distribution over all condition pairs. Then, the newly found joint probabilities were mapped into five discrete probability scores for interoperability as shown in Table \ref{tab:probability_scores}. Here, very low likelihoods of crash occuring ($P \leq 0.001$ \%) map to Score 1 and the highest risks ($P > 0.1$ \%) map to Score 5. By applying fixed thresholds to the normalized probabilities, we convert continuous values into ordinal categories that are easy to interpret and compare. This enables a clear, ordinal measure of crash likelihood that can be readily integrated into risk assessment matrices, user interfaces, or policy guidelines.

\begin{table}[ht]
\centering
\caption{Assigned Probability Scores in terms of Joint Probability Range}
\label{tab:probability_scores}
\begin{tabular}{lclc}
\toprule
\textbf{Joint Probability Range} & \textbf{Likelihood Level} & & \textbf{Probability Score} \\
\midrule
$0 \;\le\; P_{ij}\;\le\;0.010$    & Low            & & 1 \\
$0.010 \;<\; P_{ij}\;\le\;0.020$  & Medium–Low     & & 2 \\
$0.020 \;<\; P_{ij}\;\le\;0.050$  & Medium         & & 3 \\
$0.050 \;<\; P_{ij}\;\le\;0.100$  & High           & & 4 \\
$P_{ij}\;>\;0.100$                & Extreme        & & 5 \\
\bottomrule
\end{tabular}
\end{table}

\subsection{Quantification of Severity}
A severity score is a standardized metric used in risk assessment to represent the potential impact or consequence of an adverse event, independent of its likelihood. It assigns each scenario, hazard, or failure mode to a discrete level where higher values correspond to more serious outcomes. In our approach, we calculated the severity score in terms of the reduction rate by taking the difference between the posted speed of a roadway and the recommended driving speed of that roadway in cases of potential crashes due to adverse weather conditions that have effects on the road friction coefficient metric and the sight distance (i.e., visibility in term of distance).  We quantify reduction of speed as a percentage of the original speed, such that for low hazard conditions, only minimal reduction is sufficient (i.e., Severity Level 1). In contrast, in extreme hazard cases, such as black ice combined with dense fog, drivers may need to reduce their speed by more than two-thirds (Severity Level 5). 

To compute the percent reduction, at first, the safe advisory speed needs to be calculated. Safe advisory speed is the recommended maximum speed for a given roadway segment under specific environmental or traffic conditions. It is calculated to ensure drivers maintain adequate stopping distance and vehicle control when factors, such as reduced pavement friction or limited visibility, impair normal braking and maneuvering performance.  Unlike the posted speed limit, which reflects legal or design standards under ideal conditions, the safe advisory speed adjusts downward to account for real-time hazards, providing a dynamic guidance value that enhances safety by preventing drivers from traveling too fast for the prevailing conditions.

To determine safe advisory speed, we used an empirical equation \cite{Katz2012_VSL_WetWeather} derived by the Federal Highway Administration (FHWA) safety program  based on sight distance and pavement friction coefficient based on road surface conditions.

\begin{equation}
V \;=\; \frac{-3.67 \;+\;\sqrt{13.47 \;+\;\dfrac{0.12}{\mu + G}\,S}}{\dfrac{0.06}{\mu + G}}
\end{equation}
Where $\mu$ is the road surface friction coefficient value, S is the sight distance, and G is the road grade (for simplicity, we assume G = 0). This equation was modified by Nebraska-DOT; it is written as follows: 

\begin{equation}
V_s \;=\; \frac{15}{22}\,V
\end{equation}

We used a safe advisory speed equation developed for the Nebraska State Department of Transportation (DOT) \cite{NebraskaDOT2025} to determine the safe advisory speed for rural Nebraska. In this study, we considered the prevailing visibility in terms of distance as sight distance.  Also, to find the safe advisory speed and the percentage reduction to define the severity score, we need the posted speed/design speed, $V_d$. We used a minimum function to find the advisory speed, $V_a$.

\begin{equation}
V_a \;=\; \min\bigl(V_d,\,V_s\bigr)
\end{equation}

The reason we used a minimum function is that it is necessary to ensure safety by guaranteeing that the speed recommendation never exceeds what the road’s design supports, nor what current conditions allow. This conservative approach means drivers always travel no faster than the most restrictive constraint, thereby maximizing safety. In our next step, we computed the percent reduction of speed by comparing the condition-specific safe advisory speed to the baseline design speed of the roadway (i.e., 75 mph in this case). As provided in Table \ref{tab:severity_scores}, the percent-speed reduction values were defined into five levels, which indicate how severely drivers must slow down in each friction–visibility scenario. Reductions under 6.7\% (70–75 mph) indicate only a slight slowdown and are assigned severity level 1. A 6.7–20\% speed reduction (within 60–70 mph) corresponds to common moderate slowdowns, indicating a severity level 2. A 20–33.3\% speed drop (50–60 mph) captures noticeable slowdowns, is defined as severity level 3. Higher speed drops from the posted speed, i.e., 33.3–66.7\% (25–50 mph), and reductions greater than 66.7\% (less than 25 mph, which is one-third of the posted speed) are defined as severity level 4 and 5, respectively.  Lower scores indicate only modest speed adjustments typical of lightly wet or clear conditions, while higher scores flag scenarios where grip and sight distance combine to demand significant speed reduction, such as icy roads in dense fog. 

\begin{table}[ht]
\centering
\caption{Safe Advisory Speed, Reduction, and Severity Scoring}
\label{tab:severity_scores}
\begin{tabular}{lccr}
\toprule
\textbf{Safe Advisory Speed (mph)} & \textbf{Reduction (\%)} & \textbf{Severity Level} & \textbf{Score} \\
\midrule
70–75    & 0–6.67\%      & Low         & 1 \\
60–70    & 6.67–20\%     & Medium Low  & 2 \\
50–60    & 20–33.33\%    & Medium      & 3 \\
25–50    & 33.33–66.67\% & High        & 4 \\
0–25     & 66.67–100\%   & Extreme     & 5 \\
\bottomrule
\end{tabular}
\end{table}

\subsection{Risk Quantification}
The risk quantification process translates environmental conditions into a unified measure that captures both how likely a crash is and how severe its consequences would be. A probability score and corresponding severity score form the basis for our comprehensive risk model. Combining these two ordinal scores through multiplication, as shown in \ref{eq:risk}, allows for the calculation of a composite risk score. We translate these values ranging from 1 to 25 into five intuitive risk levels (Table \ref{tab:risk_levels}). Scores from 1 to 5 correspond to low risk, indicating routine conditions with minimal crash potential and impact. Scores 6–10 indicate low-medium risk, where either likelihood or severity is moderately elevated, and minor countermeasures may be warranted. Scores 11–15 fall into medium risk, highlighting scenarios that require proactive interventions, such as variable speed advisories. High risk (16–20) indicates situations with substantial probability and/or severity, and it is an ideal candidate for stricter controls like reduced speed limits or temporary road closures. Finally, Extreme risk (21–25) represents the most hazardous combinations (e.g., icy roads in dense fog), which require the strongest safety responses, including road treatments and traveler warnings. This five-tier scale transforms numeric scores into clear action thresholds for transportation practitioners.

\begin{table}[ht]
\centering
\caption{Risk Level Categories and Composite Score Ranges}
\label{tab:risk_levels}
\begin{tabular}{lc}
\toprule
\textbf{Risk Level}   & \textbf{Composite Risk Score Range} \\ 
\midrule
Low          & 1
–5   \\  
Low-Medium      & 6
–10  \\  
Medium  & 11
–15 \\  
High         & 16
–20 \\  
Extreme      & 21
–25 \\  
\bottomrule
\end{tabular}
\end{table}

\section{CASE STUDY}

\subsection{Data Generation}
In this case study, we constructed a synthetic “year-long” dataset that encompasses every possible pairing of road-surface and visibility conditions as shown in Table \ref{tab:complete_combinations}. The friction coefficient, $\mu$, is a dimensionless ratio of the tangential (friction) force to the normal load on the tire, and by definition, it cannot exceed 1. On paved highways, values above about 1 are practically unattainable. Based on our findings from the literature, as shown in Table \ref{tab:marginal_probs}, we defined four ranges in terms of road surface friction – dry (friction coefficient $\mu$   is in the range of 0.7–0.9), wet ($\mu$: 0.4–0.6), snow ($\mu$: 0.2 – 0.3), icy ($\mu$: 0.05 –0.15)\cite{FEMA2025_RoadSurfaceFriction,HPWizard2025_TireFriction,ichihara1970skid,ziadia2025weather}. We have access to a visibility sensor, i.e., VS2k-UMB \cite{Lufft2025_VS2kUMB},  in our connected and automated mobility laboratory at the University of Alabama, Tuscaloosa, AL, that has a visibility measuring range of 10 to 2,000 m, around 33 to 6,562 ft. In addition, based on our sensor reliability and findings from the literature \cite{chen2019influence,pathak20073d,Wikipedia2025_SnowClassification,astm1997274},   we partitioned the full measurement envelope into four meaningful regimes. Our “Clear” band (4,000–6,500 ft) occupies the top third of the sensor’s range, capturing essentially unrestricted daylight visibility. The “Rain/Snow” category (1,000–4,000 ft) falls in the sensor’s mid-range, where precipitation-induced scattering begins to degrade sight lines but remains measurable with good resolution. “Dense Fog” (164–1,000 ft) engages the sensor’s lower detection limits, ensuring we capture critical visibility drops under heavy mist or light fog. Finally, “Very Dense Fog” (< 164 ft) probes near the bottom of the sensor’s capability, covering extreme low-visibility conditions that approach the lower end of the instrument's visibility range. Although our sensor (VS2k-UMB) measures visibility continuously from 33 to 6,562 ft, published crash-rate studies often report visibility in coarser, sometimes nonuniform bands, such as “fog” defined anywhere from 100 to 1,000 ft or “rain” lumped with any visibility below 2,000 ft. As a result, the crash-rate categories based on visibility ranges in the literature from table \ref{tab:marginal_probs} do not align exactly with our four ranges.  By defining our bands to match the VS2k’s measurement envelope and by splitting the lower end into two ranges,  we capture the rapid increase in crash rates seen below about 200 ft, something that is missing from various coarser literature. 
Thus, using the 16 friction–visibility combination from table \ref{tab:complete_combinations}, we sampled 100 friction coefficients and 100 sight‐distance values from truncated normal distributions clipped to their respective bounds, yielding 1,600 unique pairs of friction $\mu$ and sight distance S. The normal distribution was used because real world measurements tend to cluster around typical condition values with symmetric variability due to factors like surface texture heterogeneity or gradual changes in atmospheric clarity. The normal distribution \cite{eugene2002beta} naturally captures this type of ``around the mean'' behavior and accords with the Central Limit Theorem \cite{feller1971introduction}, which describes the behavior of many small, independent effects, such as road wear, temperature fluctuations, and particle scattering in air, that combine to produce a bell-shaped spread. In the equation \ref{normal}, $f(x,\mu_n,\sigma) $is the probability density function of a normal distribution with mean $\mu_n$ and standard deviation 
$\sigma$. In this study, we sample friction coefficients and visibility distances from this distribution and then truncate at the condition bounds to realistically model natural variability around the mean values.

\begin{equation}
f(x;\mu_n,\sigma) \;=\; \frac{1}{\sigma\sqrt{2\pi}}
\exp\!\Biggl(-\frac{(x-\mu_n)^{2}}{2\sigma^{2}}\Biggr)
\label{normal}
\end{equation}\\

\begin{figure}[!htbp]
  \centering
  \includegraphics[width=0.8\textwidth]{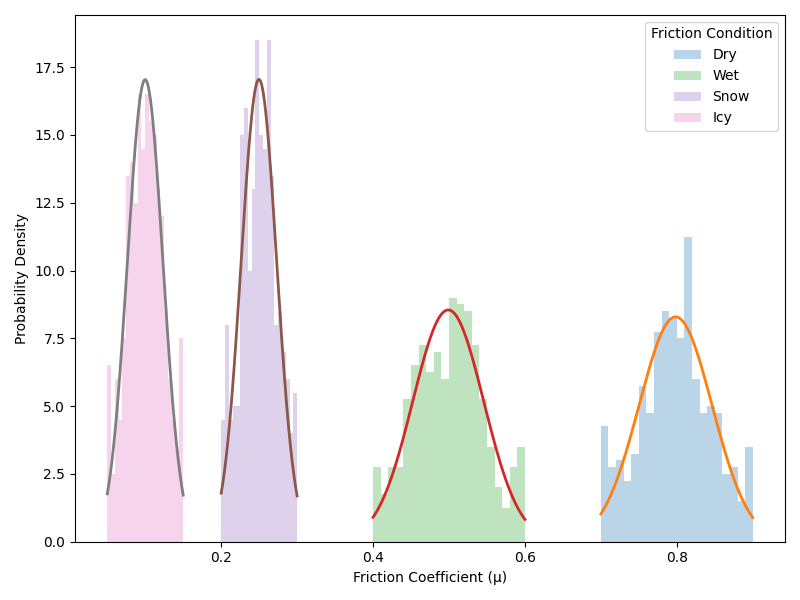}
  \caption{ Overlayed Friction Distributions with Fitted Normal Curves).
  }
  \label{fig:friction_dist}
\end{figure}

\begin{figure}[!htbp]
  \centering
  \includegraphics[width=0.8\textwidth]{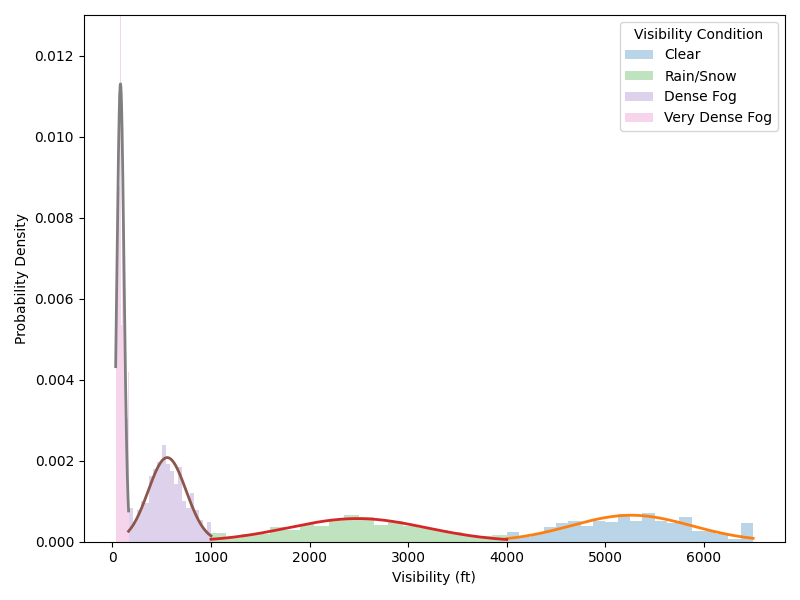}
  \caption{ Overlayed Visibility Distributions with Fitted Normal Curves }
  \label{fig:visibility_dist}
\end{figure}

Having defined our truncated normal sampling framework with Equation \ref{normal}, we next verify that the synthetic data adhere to the intended statistical model. Figure \ref{fig:friction_dist} overlays the normalized histograms and fitted Gaussian curves for the four friction regimes—Dry, Wet, Snow, and Icy. Each distribution peaks near its midpoint (e.g., $\mu \approx$0.80 for Dry, $\mu \approx$ 0.10 for Icy) and shows minimal overlap, confirming that our sampling faithfully captures both central tendency and variability within the specified bounds. 

In Figure \ref{fig:visibility_dist}, we apply the same approach to visibility, plotting the four bands of Clear, Rain/Snow, Dense Fog, and Very Dense Fog on a common density scale; here, the y axis limited to 0.013 to fully display the tallest curves fitted under the bell curve. The sharper, taller peaks of the low visibility regimes contrast with the broader, flatter profiles of high-visibility conditions, illustrating that our truncated normals reproduce the sensor’s measurement envelope of 33–6562 ft with appropriate dispersion. Both of these visualizations demonstrate that our data generation process produces statistically robust samples for both friction and visibility, laying a solid foundation for the subsequent risk scoring analysis.

\subsection{Results and Discussion}
\begin{figure}[!htbp]
  \centering
  \includegraphics[angle=90,width=0.55\textwidth]{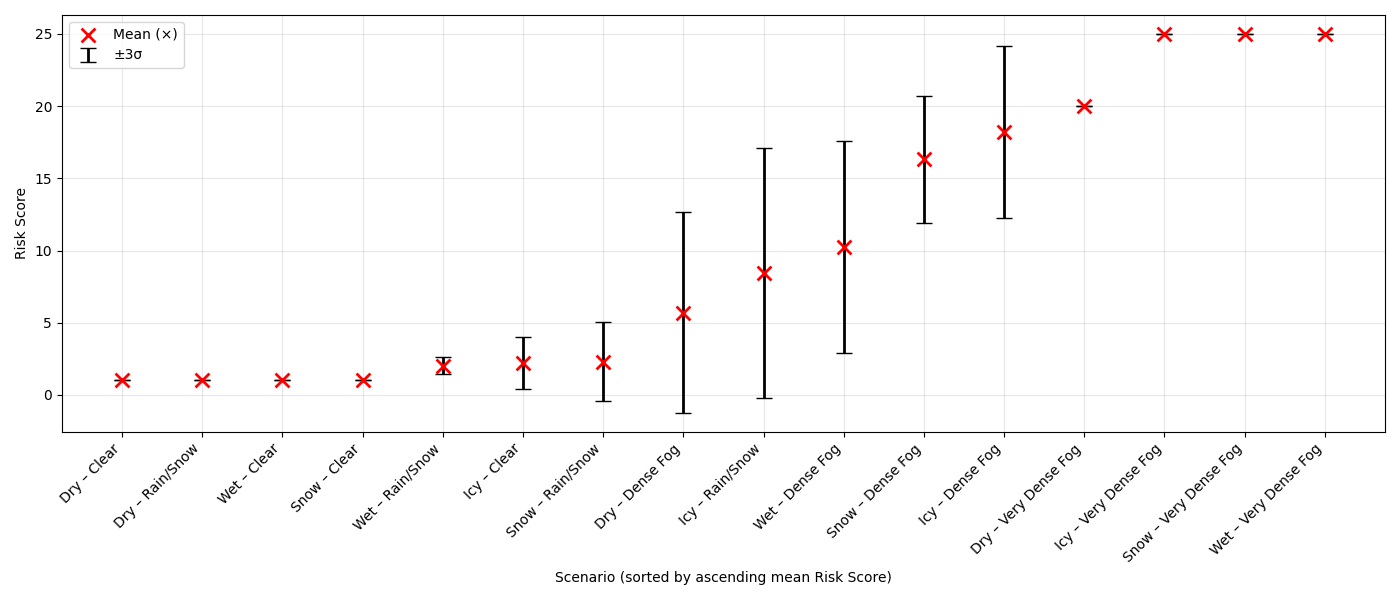}
  \caption{ Risk Score Distributions Ordered by Mean (x and $\pm 3\sigma$) }
  \label{fig:scoreing}
\end{figure}

Based on our synthetic dataset of 1,600 cases (16 friction–visibility combinations × 100 samples each), we computed a composite risk score for every sample by multiplying its probability score (derived from normalized joint crash probabilities) with the corresponding severity score (based on percentage speed reduction). Risk scores thus range from 1 (minimal likelihood and impact) up to 25 (highest crash probability paired with maximum speed‐reduction severity), as provided in Table \ref{tab:risk_levels}. Figure \ref{fig:scoreing} presents each scenario’s full Risk Score distribution (blue markers connected) alongside the scenario mean (red “x”) and $\pm3\sigma$ error bars ordered by ascending value of mean. This figure reveals a clear, monotonic escalation as conditions worsen. 

The first six scenarios (Dry–Clear, Dry–Rain/Snow, Wet–Clear, Snow–Clear, Wet–Rain/Snow, Icy–Clear, Snow–Rain/Snow) all fall at or below a mean score of 5 with negligible spread. Here, both crash probability (high friction and/or clear line-of-sight) and required speed reductions remain minimal, so every one of the 100 samples in each scenario clusters at the floor of the risk scale. These conditions correspond to everyday driving with only routine advisory signage needed in practice. However, even though the weather conditions in these four scenarios may seem benign, a low risk value is still assigned to them as human error always remains; this is why a risk score of 1 is assigned here. 

Moving into the next two conditions, i.e., Dry–Dense Fog (mean $\approx$ 5.6) and Icy–Rain/Snow (mean $\approx$ 8.4), rise into the 6–10 window. Meaning that in Dry–Dense Fog condition, sufficient pavement grip still lets drivers feel in control, but the thick fog cuts visibility so much that recommended speeds must drop by 20–33\%. In Icy–Rain/Snow, slippery ice greatly increases crash risk, but visibility isn’t as badly affected, so speed reductions stay in a lower range. Because our scoring uses discrete bins, every one of the 100 samples in each scenario lands in the same medium-risk category without crossing a boundary, demonstrating that these thresholds align with realistic use cases.

As fog thickens over wet or icy roads (Wet–Dense Fog, Snow–Dense Fog), mean risk scores shift into the Medium–High band (11–15), and under Icy–Dense Fog, they enter the High band (16–20). These scenarios also exhibit larger $\pm 3 \sigma$ bars, indicating small changes in visibility or road friction can lead to bigger swings in the recommended safe speed. This wider spread validates that our scoring captures the inherent uncertainty of moderate-to-severe conditions. In a real-world scenario, that means traffic agencies should step up warnings using electronic message boards, stricter speed limits, or even closing lanes, so that drivers slow down enough and stay safe in these tougher conditions.

Under Very Dense Fog (visibility < 164 ft), most surface conditions based on weather like Wet (Rain), Snow, and Icy both crash probability and severity to their maximum bins. However, on Dry pavement, the mean risk score hovers around 20 rather than 25. This occurs because, despite near-zero sight distance, a perfect friction value (Dry in this case) still slightly mitigates crash likelihood and required speed reduction, preventing the severity component from reaching its maximum value.  The smaller $\pm 3 \sigma$ spread under Dry–Very Dense Fog further confirms that high friction stabilizes risk outcomes even in the worst visibility.

\begin{figure}[!htbp]
  \centering
  \includegraphics[width=0.75\textwidth]{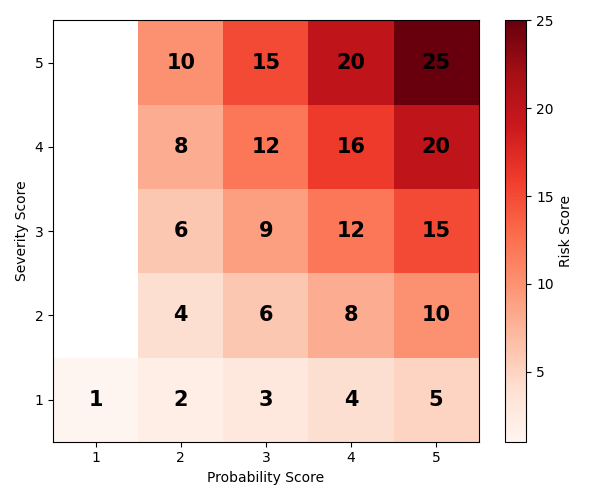}
  \caption{ Heatmap of composite Risk Scores (Probability × Severity)}
  \label{fig:heatmap}
\end{figure}

Figure \ref{fig:heatmap} visualizes the composite Risk Score as the product of discrete Probability Scores (x-axis) and Severity Scores (y-axis). Each cell’s color intensity and overlaid number represent the resulting Risk Score, ranging from 1 (lightest, bottom-left) to 25 (darkest, top-right). ells in the lower-left corner (low probability, low severity) fall into the Low risk band (1–5), while those in the upper-right reach the Extreme band (21–25). Intermediate cells trace out the Medium (6–10), Medium–High (11–15), and High (16–20) bands, providing an intuitive map for how different environmental and operational conditions translate into actionable risk levels.

By distinguishing Dry from the other surfaces in this extreme‐fog regime, our scoring framework captures the real-world subtlety that good pavement grip offers a modest safety buffer when visibility collapses, while still treating all Very Dense Fog scenarios as critical, falling into the “High” to “Extreme” thresholds so agencies can calibrate interventions appropriately.

\section{CONCLUSIONS}

In this study, we present a comprehensive roadway hazard risk assessment framework designed to systematically quantify both the probability and severity of crash occurrences associated with specific roadway hazards. The framework integrates factors, such as visibility and surface condition, enabling a structured evaluation of risk under a wide range of scenarios. To assess the effectiveness of the proposed approach, we developed a synthetic dataset representing a full calendar year, comprising every possible combination of road surface (e.g., dry, wet, icy) and visibility conditions (e.g., clear, light fog, dense fog). This exhaustive dataset allows for a controlled analysis of hazard interactions and their impact on crash risk.

The results of our case study demonstrate that the combined Probability-Severity scoring method generates a consistent and interpretable risk gradient across all examined conditions. Scenarios characterized by minimal hazard exposure such as dry pavement under clear visibility consistently yield low and stable risk scores. Conversely, conditions involving extreme visibility impairment, such as dense fog, result in uniformly high risk levels, regardless of surface type. Of particular interest are the intermediate regimes, especially those combining dense fog with wet or icy road surfaces. These cases not only produce higher average risk scores but also exhibit significant variability, underscoring the complexity and unpredictability of crash risk under compound hazard scenarios.

These findings validate the utility of the proposed framework in capturing nuanced risk variations and highlight its potential for supporting real time safety applications. In particular, the ability to identify transitional or threshold conditions where risk sharply escalates can inform the deployment of adaptive speed advisories, hazard warnings, and other proactive interventions. As such, this work lays a foundational basis for translating probabilistic risk assessment into realistic roadway safety strategies suitable for dynamic and hazard-prone driving environments.




\section{Acknowledgments}
This study was financially supported by the Federal Motor Carrier Safety Administration (FMCSA) of the United States Department of Transportation (US DOT). The views expressed in this paper are solely those of the authors, who are responsible for the accuracy and factual content presented. The contents do not necessarily reflect the official views or policies of the FMCSA and US DOT.

We used ChatGPT to help rephrase parts of our own writing to improve clarity.

\section{Authors Contribution}
\textbf{Suhala Rabab Saba and Sagar Dasgupta:} conceptualization, methodology, data generation, data analysis, and writing – original draft; \textbf{Mizanur Rahman:} conceptualization, methodology, writing – original draft, and funding acquisition;   \textbf{Li Zhao, Nathan Huynh, Mehmet Can Vuran, Qiang Liu and Eren Erman Ozguven:} conceptualization, writing – review and editing, and funding acquisition.

\section{CONFLICT OF INTEREST}
The authors declare no conflicts of interest with any other entities or researchers.

\newpage
\newpage

\bibliographystyle{trb}
\bibliography{trb_template}
\end{document}